\documentclass[aps,prc,twocolumn,nofootinbib,showpacs,preprintnumbers,amsmath,amssymb]{revtex4}

\usepackage{graphicx}
\usepackage{graphicx}
\usepackage{dcolumn}
\usepackage{bm}
\newcommand{\be}{\begin{eqnarray}}
\newcommand{\ee}{\end{eqnarray}}

 \newcommand{\gsim}{\mathrel{\hbox{\rlap{\lower.55ex \hbox {$\sim$}}
                   \kern-.3em \raise.4ex \hbox{$>$}}}}
\newcommand{\lsim}{\mathrel{\hbox{\rlap{\lower.55ex \hbox {$\sim$}}
                   \kern-.3em \raise.4ex \hbox{$<$}}}}

\begin{document}

\title{Fluidity and supercriticality of the QCD matter \\
created
in relativistic heavy ion collisions}

\author{Jinfeng Liao}\email{jliao@lbl.gov}

\author{Volker Koch}\email{vkoch@lbl.gov}

\address{Nuclear Science Division, Lawrence Berkeley National Laboratory, MS70R0319, 1 Cyclotron Road, Berkeley, CA 94720.}




\begin{abstract}
 In this paper we discuss the fluidity of the hot and dense QCD
matter created in ultrarelativistic heavy ion collisions in
comparison with various other fluids, and in particular suggest
its possible supercriticality. After examining the proper way to
compare non-relativistic and relativistic fluids from both
thermodynamic and hydrodynamic perspectives, we propose a new
fluidity measure which shows certain universality for a remarkable
diversity of critical fluids. We then demonstrate that a fluid in
its supercritical regime has its fluidity considerably enhanced.
This may suggest a possible relationship between the seemingly
good fluidity of the QCD matter produced in heavy ion collisions
at center of mass energy of $\sqrt{s}=200 \, \rm AGeV$ and the
supercriticality of this matter with respect to the
Critical-End-Point on the QCD phase diagram. Based on such
observation, we predict an even better fluidity of the matter to
be created in heavy ion collisions at LHC energy and the loss of
good fluidity at certain lower beam energy. Finally based on our
criteria, we analyze the suitability of a hydrodynamic description
for the fireball evolution in heavy ion collisions at various
energies.\\
\end{abstract}


\maketitle

\vspace{0.5in}

\section{Introduction}

The exploration of the QCD phase diagram as well as the
quantitative characterization of QCD matter is one of the most
interesting challenges and questions in strong interaction
physics. Hot and dense QCD matter can be created in the laboratory
by means of heavy ion collisions, and experiments at the
CERN Super Proton Synchrotron (SPS)
and at the  Relativistic Heavy Ion Collider (RHIC) have, over the years,
revealed many intriguing and
unexpected properties of this matter. For example the measurement
of an unexpectedly large elliptic anisotropy $v_2$
\cite{Voloshin:2008dg} can be reproduced within the framework of
hydrodynamics \cite{hydro_review}, at least for low transverse
momenta ($p_t$). This observation has led to the conjecture, that
the matter produced in these collisions is strongly interacting,
with nearly ideal fluidity \cite{Schaefer:2009dj}. Several
microscopic explanations for this behavior have been suggested
\cite{Liao:2006ry}\cite{Carsten}\cite{Pisarski}, for example the
``magnetic scenario'' which features the co-existing electric and
magnetic components of the plasma with the magnetic one ultimately
enforcing the QCD confinement transition.

Arguments for the nearly ideal fluidity at center of mass energies of
$\sqrt{s}=200 \, \rm AGeV$, quantitatively
represented by a rather small ratio of shear-viscosity over
entropy-density, $\eta/s$, have come from different directions.
Firstly, the $v_2$ data measured at RHIC have provided rather
stringent constraints on viscous hydrodynamic calculations: the
current status based on this approach is that an {\em upper limit}
$\eta/s \le 6/(4\pi)$ may be set \cite{hydro_review}.
 (Some caveats on interpretation of $v_2$ data
however should be kept in mind, such as hidden non-collective
contributions to $v_2$ \cite{Liao:2009ni} or alternative mechanism
of producing $v_2$\cite{Koch:2009wk}). Secondly, recent
developments from the AdS/CFT correspondence \cite{Policastro:2001yc}
have hinted at a universal {\em lower bound} $\eta/s\ge 1/(4\pi)$.
While in principle violating examples have been found (see
detailed discussions in e.g. \cite{Cherman:2007fj}), for all
practical purposes this value serves as a useful benchmark for good
fluidity of {\em relativistic fluids}. Finally, the fluidity of
the QCD fluid has been compared \cite{Csernai:2006zz,Lacey:2006bc}
to a few commonly known substances like Helium-4, water, nitrogen etc,
using the measure $\eta/s$: it has been concluded that the QCD
fluid has smaller $\eta/s$ as compared with all those normal
substances.

In this paper we will revisit the concept of fluidity. For a
sensible  discussion of fluidity one needs to clearly distinguish
the concept of \emph{fluidity} from the concept of applicability
and validity of hydrodynamics. The concept of fluidity of a
substance can only be meaningful if the fluidity is defined
exclusively in terms of properties of the substance itself. The
question of applicability of hydrodynamics, on the other hand,
requires the effective mean free path\footnote{To simplify the
argument, in the rest of the introduction we will use the familiar
concepts of ``mean free path'' and particles, noting that these
are limited to a kinetic description of a sufficiently dilute
system. In section \ref{sec:section_3} we will derive a more
general expression for the relevant length scale which will not
require the applicability of kinetic theory. In the kinetic limit
it reduces to the mean free path. } to be small compared to the
size of the system or rather the typical wavelengths of the
excitations which are supposed to be described within
hydrodynamics. Consequently, in this case we compare a property of
the system, the mean free path, with an \emph{external} scale
characterizing the situation we want to describe within
hydrodynamics, for example  a sound mode of a given wavelength.
And given a sufficiently large external scale, hydrodynamics is
always applicable. For example we observe sound propagation in
both water and air, and yet one would be inclined to assign a
better \emph{fluidity} to water than to air. Therefore a
meaningful definition of \emph{fluidity} needs to measure the
effective mean free path in terms of a length scale inherent to
the substance itself. One obvious choice is the interparticle
distance, or more generally, the density-density correlation
length. This allows the comparison of substances with vastly
different inherent length scales, such as water vs. a Quark-Gluon
Plasma. Obviously, the minimum wavelength (in meters) for sound
propagation even in a weakly interacting QGP is still considerably
shorter than that of a strongly interacting atomic or molecular
substance. However, if we ask the question ``what is the minimal
wavelength for sound propagation measured in units of the
interparticle distance?'' then a comparison between these
substances becomes meaningful, and the strongly interacting
substance will likely exhibit a better fluidity.

Following these general considerations, in this paper we discuss
the fluidity of the hot and dense QCD matter created at RHIC by
comparing it with normal non-relativistic fluids and by applying
valuable insights from those fluids. While such a comparison is
not new (see e.g. \cite{Csernai:2006zz}\cite{Lacey:2006bc}), we
differ from all previous approaches in a few distinct points.
First of all, we carefully examine the difference between the
relativistic and non-relativistic fluids. This includes their
different inertia: the relativistic inertia, i.e. the enthalpy, reduces to
 the mass density in the non-relativistic regime. This
also includes the choice of reference scale, as discussed above:
while the temperature may serve as a reasonable estimator for the
interparticle distance for relativistic fluids, it certainly does
not in the non-relativistic limit. Based on these considerations,
we propose a new measure of fluidity for comparing various fluids.
Furthermore we emphasize, for the first time, the possible
relevance of the so-called supercritical fluid for the matter
produced at RHIC and discuss important implications for the
expected fluidity for the matter produced in heavy ion collisions
at different center of mass energies. We will also elaborate on
potential consequences for the search of the QCD
Critical-End-Point via a beam energy scan.

As just discussed, a closely related, but different, question is
the applicability of hydrodynamics in heavy ion collisions at
various beam energy $\sqrt{s}$. For the description of the
dynamical evolution of system within hydrodynamics to be valid the
length scale characterizing the variations of the system needs to
be large compared to the effective mean free path of the (quasi)
particles within the fluid \cite{hydro_review}\cite{Betz:2008me}.
Thus,  good fluidity of the underlying matter may not guarantee
applicability of hydrodynamics, if the typical gradients of the
flow field are large, e.g. when the system size is very small.
Based on an analysis of sound wave attenuation, we will provide
quantitative criteria for the applicability of hydrodynamics, and
evaluate the situations at SPS, RHIC, and LHC respectively.

This paper is organized as follows: In Section \ref{sec:section_2}
we will discuss the difference and relation between a
non-relativistic(NR) fluid and a relativistic(R) fluid. In Section
\ref{sec:section_3} we will then propose a new fluidity measure,
which is applicable for both relativistic and non-relativistic
systems. Based on this new measure we will compare various fluid
systems and demonstrate the improvement of fluidity in a fluid's
 supercritical regime. In
Section \ref{sec:section_4} we will use Lattice QCD results to
construct equal-pressure lines on the QCD phase diagram. We then
discuss the relationship between fluidity and supercriticality for
heavy ion collisions, the evolution of fluidity with beam
energies, and its implications for the search of the QCD
Critical-End-Point(CEP). Finally in Section \ref{sec:section_5}
the applicability of hydrodynamics in heavy ion collisions at
various energies will be discussed.

\section{Relativistic and Non-Relativistic Fluids}
\label{sec:section_2}

When comparing a relativistic (R) with a non-relativistic (NR)
fluid, one needs to carefully keep track of the mass terms, which
customarily are neglected in non-relativistic thermodynamics. In
non-relativistic thermodynamics, the basic thermodynamic relation
\begin{eqnarray}
   E_{NR}=T S - p V + \mu_{NR}\, N
\end{eqnarray}
does not take into account the mass of the
particles. Here $E_{NR}$ refers to the {\em kinetic and interaction} energy
of the particles. Similarly, the chemical potential, which represents the
increase of energy  by the addition of one extra particle, does
not account for the particle's mass. The relativistic version of
the basic thermodynamic relation,
\begin{eqnarray}
   E_R=TS - pV + \mu_R N,
\end{eqnarray}
on the other hand, takes into account the particle masses. Its non-relativistic
limit can be obtained by simply including the mass terms
in both the energy and the chemical potential
\begin{eqnarray}
   E_R=E_{NR} + m
\\
   \mu_R=\mu_{NR}+m .
\end{eqnarray}
As we will discuss below (see also \cite{Landau_book,Weinberg_book}), the
thermodynamic quantity
entering hydrodynamics is the enthalpy density, $w$ defined as
\begin{eqnarray}
   w_R=\epsilon + p = T s + \mu_R \, n
\end{eqnarray}
where $\epsilon$ is the energy density, $p$ the pressure, $s$ the
entropy-density, and $n$ is the particle-density. As we will show, the
non-relativistic limit of
hydrodynamics involves the non-relativistic limit of the enthalpy,
$w_R$, \emph{including} the mass term
\begin{eqnarray}
   w = Ts + (\mu_{NR} + m) n \stackrel{T \ll m}{\rightarrow}  m \,n
   \equiv \rho
\end{eqnarray}
and it is dominated by the mass density. In the ultra-relativistic
limit $T\gg \mu_R$, on the other hand, the enthalpy density is given
by
\begin{eqnarray} \label{eqn_thermo_ts}
   w \stackrel{T \gg \mu_R}{\rightarrow} Ts.
\end{eqnarray}
The kinematic viscosity, which is defined as the ratio of the shear viscosity over the enthalpy density
\begin{eqnarray}
\nu=\frac{\eta}{w}
\end{eqnarray}
usually serves as a measure for dissipation \cite{Landau_book}. While the widely used
ratio of shear-viscosity over entropy density, $\eta/s$ is indeed
related with the kinematic viscosity for  ultra-relativistic
systems, it misses the dominant mass term for a non-relativistic
fluid.

In the following, we further demonstrate this point both in
thermodynamics and in hydrodynamics\footnote{ To better keep
track of the mass terms, in the following two subsections we make
explicit the dependence on the speed of light $c$, the Boltzmann
constant $k_B$ and the Planck constant $\hbar$, while in the rest
of the paper we use natural units with these constants taken to be
unity.}.

\subsection{Thermodynamics}

We start with the example of a classical (Boltzmann), relativistic
free gas at temperature $T$ and {\em fixed (net-)particle density}
$n=n_P-n_{\bar P}$. Following standard statistical mechanics in
e.g. \cite{Huang_book}, the Helmholtz free energy
density is given by
\begin{eqnarray}
 f(T,n) =&&  n\, (k_B T)\, {\bigg\{} - \sqrt{1+\left( \frac{2\, I\left[\beta
mc^2\right]}{n\tilde{\lambda}^3} \right)^2} \nonumber \\
&&+ \mathbf{ln}\left[\frac{\frac{n\tilde{\lambda}^3}{I\left[\beta
mc^2\right]}+\sqrt{(\frac{n\tilde{\lambda}^3}{I\left[\beta
mc^2\right]})^2+4}}{2} \right] {\bigg\}} \quad
\end{eqnarray}
with $\beta\equiv 1/(k_B T)$ and $\tilde{\lambda}=(2\pi^2)^{1/3}
(\beta \hbar c)$. Here we have introduced
the re-scaled momentum integral,  $\tilde{p}=\beta p\, c\,$,
\begin{equation}
 I[\beta mc^2] \equiv \int_0^\infty d\tilde{p}\,
\tilde{p}^2\, e^{-\sqrt{\tilde{p}^2+(\beta mc^2)^2}}.
\label{eq:momentum_integral}
\end{equation}
Using the usual thermodynamic relations we obtain expressions for other quantities, such as
the chemical potential and the energy density:
\begin{eqnarray}
\mu=\frac{\partial f}{\partial n}=(k_B T)\,
\mathbf{ln}\left[\frac{\frac{n\tilde{\lambda}^3}{I\left[\beta
mc^2\right]}+\sqrt{(\frac{n\tilde{\lambda}^3}{I\left[\beta
mc^2\right]})^2+4}}{2} \right]
\end{eqnarray}
\begin{eqnarray}
\epsilon &=& f - T\frac{\partial f}{\partial T} \nonumber \\
&=& n\, (k_B T) \, \sqrt{1+\left( \frac{2\, I}{n\tilde{\lambda}^3}
\right)^2} \, \left(3- (\beta mc^2)\, \cdot \frac{I'}{I}  \right)
\end{eqnarray}

Next we examine the non-relativistic  limit by taking $\beta
mc^2\to \infty$ in function $I[y]$, Eq.\ref{eq:momentum_integral},
(and its derivative $I'\equiv d I[y] / d y$):
\begin{eqnarray} \label{eqn_thermodynamics}
&& f \to \,\, NR : \, nmc^2 + n (k_B T)
\left[\mathbf{ln}(n\lambda^3)-1 \right] \nonumber \\
&& \mu \to \,\, NR : \, mc^2 + (k_B T) \mathbf{ln}(n\lambda^3)
\nonumber \\
&& \epsilon \to \,\, NR : \, n mc^2 + \frac{3}{2} n (k_B T)
\end{eqnarray}
with $\lambda\equiv (2\pi\hbar^2/mk_B T)^{1/2}$. Obviously,  in
the non-relativistic  regime the energy associated with the rest
mass dominates the chemical potential, the free energy density,
the energy density, and the enthalpy density. Contrary to that,
neither  entropy density $s=(\epsilon-f)/T$  nor the pressure
$p=\mu n -f$, have an explicit dependence of the mass term, as
intuitively expected.

\subsection{Hydrodynamics}

We now turn to the hydrodynamics in the relativistic  and
non-relativistic regime. Since the non-relativistic limit of
relativistic hydrodynamics is discussed in textbooks
\cite{Landau_book,Weinberg_book}, we will be brief here and just
remind ourselves of the essential points, in particular how the
relativistic inertia i.e. the enthalpy density $w$ is replaced in
the non-relativistic limit by the mass density $\rho c^2$. We
start with the non-relativistic  Navier-Stokes (N-S) equation\footnote{In this
discussion of N-S equation we neglect bulk
viscosity and assume constant shear viscosity across the fluid.}:
\begin{eqnarray}
[\partial_t + \vec v \cdot \vec \bigtriangledown] \vec v = -
\frac{\vec \bigtriangledown\, p}{\rho} + \frac{\eta}{\rho}
\vec{\bigtriangledown_j \Sigma^{j i}}
\end{eqnarray}
with the non-relativistic shear tensor $\Sigma^{ji}=\partial_j v_i
+\partial_i v_j-\frac{2}{3} \delta_{ji}\vec
\bigtriangledown\cdot\vec v$. The corresponding relativistic N-S
equation is given by\footnote{It is well-known that for
relativistic hydrodynamics the derivative expansion to only first
order, i.e. the N-S form, has causality problem and a consistent
treatment requires higher orders in derivative expansion, see e.g.
\cite{hydro_review}.}:
\begin{eqnarray}
\gamma^2  [\partial_t + \vec v \cdot \vec \bigtriangledown] \vec v
=&& - \frac{1}{w/c^2} [\vec \bigtriangledown\, p + \frac{\vec
v}{c}
\partial_0 p] \nonumber \\
&&  + \frac{\eta}{w/c^2} \vec{\partial_\nu \Sigma^{\nu i} } \,\,
\end{eqnarray}
with the relativistic shear tensor $\Sigma_{\mu\nu}=c\,
[\partial_\mu u_\nu + \partial_\nu u_\mu -(u\cdot
\partial)u_\mu u_\nu + \frac{2}{3}(u_\mu u_\nu - g_{\mu \nu})(\partial\cdot
u)]$, $\gamma=1/\sqrt{1-v^2/c^2}$, $u_\mu=\gamma(1,\vec v/c)$, and
$\partial_0 =\frac{1}{c}
\partial_t $. In the non-relativistic limit one has $\gamma \to 1$, and $w/c^2 \to \rho$ and thus in
leading order of $v/c$ recovers the non-relativistic Navier-Stokes
equation \cite{Landau_book}.

To further elaborate the point, let us consider the propagation
and attenuation of a sound wave  of frequency $\omega$ and wave
vector $k=2\pi/\lambda_s$ ($\lambda_s$ is the wavelength ) in the
 presence of dissipation, characterized by the shear viscosity
$\eta$. The dispersion relation for the wave is given by
\begin{eqnarray} \label{eqn_sound_dispersion}
\omega= c_s\, k - \frac{i}{2}\, k^2\times  \left\{\begin{array}{c}
\frac{\frac{4}{3}\eta}{w/c^2}\, ,\quad
\mathrm{R\,\, fluid} \\
\frac{\frac{4}{3}\eta}{\rho}\, ,\quad \mathrm{NR\,\, fluid}
\end{array}\right\}
\end{eqnarray}

To take into account possible bulk viscosity $\zeta$, one simply
makes the replacement $\frac{4}{3}\eta \to \frac{4}{3}\eta+\zeta$.
Furthermore the speed of sound $c_s$ is given by
\begin{eqnarray}
c_s = \left\{\begin{array}{c} \sqrt{\frac{\partial P}{\partial (\epsilon/c^2)}}
\, , \quad  \mathrm{R\,\, fluid} \\
\sqrt{\frac{\partial P}{\partial \rho}}\, , \quad \mathrm{NR\,\,
fluid}
\end{array}\right\}
\end{eqnarray}
As one can see, the same correspondence $(w,\epsilon)_R \to (\rho
c^2)_{NR}$ appears again as in the thermodynamics
Eq.(\ref{eqn_thermodynamics}).

\subsection{Discussion on $\eta/s$}

We end this section by a discussion of the ratio of
shear-viscosity over entropy-density-ratio, $\eta/s$, which has
attracted a considerable interest in various fields of physics.

We first recall, from the perspective motivated by studying the
QGP via heavy ion collisions, why $\eta/s$ is a useful measure for
the relativistic fluid, as was first pointed out in the seminal
paper \cite{Danielewicz:1984ww}. Consider a Bjorken-type
longitudinal expansion of the quark-gluon plasma (with low
baryonic density) formed in heavy ion collisions: its system size
is limited by $c \tau$ at early time. With the presence of shear
viscosity $\eta$, there will be dissipative effect (e.g. entropy
generation) characterized by the ratio $\frac{\eta/w}{\tau}$
(roughly the Knudsen number) which, by thermodynamic relation
Eq.(\ref{eqn_thermo_ts}) $w \simeq Ts$, leads to $\frac{\eta}{s}\,
\frac{1}{T\tau}$. To ensure a controllable dissipative correction
to the ideal hydrodynamics, one requires firstly $\eta/s$ is small
enough and, secondly, $T\tau$ is of the order 1 or bigger. The
former shall be a property of the underlying QGP while the latter
means the viscous effect is most severe at early time and the
hydrodynamic evolution may not be  a good approximation at too
early time. (Note in this discussion we adopted natural
relativistic units with $c=1$). One can draw two conclusions from
this discussion: (a) $\eta/s$ can serve as a good measure of
fluidity for a relativistic fluid and the smaller it is the better
the fluidity; (b) the ability of $\eta/s$ to serve such a role is
actually inherited from $\eta/w$.

The discussion above, however, leads to the observation that {\em $\eta/s$
for a non-relativistic fluid does NOT necessarily provide a good
measure for its fluidity}, because the role of the relativistic
$\eta/w$ corresponds to the non-relativistic $\eta/\rho$. This is
certainly not a new lesson: it has been known from Navier and
Stokes's time that what really matters for usual non-relativistic
fluids' fluidity is not the dynamical viscosity $\eta$ itself but
rather the so-called kinematic viscosity $\eta/\rho$ (see e.g.
\cite{Landau_book}). Indeed by looking at actual data for water in the
liquid and vapor phase at the same pressure one finds that liquid
water has about one order of magnitude bigger shear viscosity,
$\eta$, but nonetheless ``wins the fluidity contest'' since its
kinematic viscosity, $\eta/\rho$, is about two orders of magnitude
smaller than that of the vapor phase. We note that, the expression
$\eta/w$ for a relativistic fluid is a relativistic version of the
kinematic viscosity (see also related discussions in
\cite{Schaefer:2009dj}). To conclude, we emphasize that for a
non-relativistic fluid $\eta/\rho$ is different from $\eta/s$ (as
is evident from the thermodynamics discussion) and only
$\eta/\rho$ serves as a good measure of fluidity (as is evident
from the hydrodynamics discussion). This also implies that using
$\eta/s$ to compare the fluidity between a relativistic fluid
(like the QGP, the AdS/CFT plasma) and a non-relativistic fluid
(like Helium, water or cold Fermion gas \cite{Schaefer:2009kj} )
may not be as informative as one would expect.

\section{A New Fluidity Measure}
\label{sec:section_3}

In this section we propose a new fluidity measure that is suitable
for comparison between relativistic  and non-relativistic  fluids.
In the following we will analyze the propagation of sound modes in
the presence of dissipation (viscosity) and ask ourselves under
what condition the dissipative (viscous) terms in the equations
prevent sound from propagating. The advantage of this strategy is
that we stay entirely within the framework of viscous
hydrodynamics and do not need to make additional assumptions, such
as the applicability of kinetic theory. As a consequence our
fluidity measure will be expressed in terms of well defined
quantities such as the shear viscosity, the speed of sound, etc.

We start with the sound dispersion relation in
Eq.(\ref{eqn_sound_dispersion}). By requiring  the imaginary
part of the frequency, ${\mathcal Im}\omega$, to be small in magnitude as
compared
with its real part ${\mathcal Re}\omega$,  we obtain:
\begin{eqnarray}
|\frac{{\mathcal Im}\omega}{{\mathcal Re}\omega}| <<1 && \to \nonumber \\
\lambda_s &=& \frac{2\pi}{k} >> \frac{4\pi}{3} L_{\eta}  \\
L_{\eta} & \equiv & \left\{\begin{array}{c} \frac{\eta}{(w/c^2)\,
c_s}\, ,\quad
\mathrm{R\,\, fluid} \\
\frac{\eta}{\rho\, c_s}\, ,\quad \mathrm{NR\,\, fluid}
\end{array}\right\}
\end{eqnarray}
The above equation implies that if a sound wave has its wavelength
$\lambda_s$ comparable (or even smaller) than $L_\eta$, it will be
quickly damped on (or shorter than) a time scale of its period and
spatially on a length scale about (or shorter than) its
wavelength, which essentially means it can not propagate away in
the medium. Therefore, the physical meaning of the length $L_\eta$
introduced above is to provide a measure for the {\em
minimal} wavelength of a sound wave to propagate in such a viscous
fluid\footnote{We note that other types of dissipative processes
like
 thermal conduction may also be present and thus introduce
different length scales. For example when the sound wave period
$\tau_s \sim 1/(c_s k)$ becomes larger than the thermal relaxation
time scale $\tau_T \sim 1/(D_T k^2)$ (with $D_T$ the thermal
diffusivity) at wavelength smaller than $D_T/c_s$ , then heat
transport becomes rather efficient and the sound propagation
becomes isothermal (see examples in e.g. \cite{supercritical}).
Nevertheless additional sources for dissipation do not change the fact that
the ``good'' sound modes shall have their wavelengths (at least)
larger that the $L_\eta$ set by shear viscosity only. }. A more
quantitative criteria will be discussed in Section
\ref{sec:section_5} and given in Eq.(\ref{eqn_criteria}).

Furthermore the length $L_\eta$ has the meaning of an effective
mean-free-path (MFP) in terms of microscopic fluid particle motion.
This becomes transparent in a weakly coupled gas: taking the
non-relativistic gas as an example, the shear viscosity according
to kinetic transport is
\begin{equation}
\eta\sim \rho v_T l_{MFP}
\end{equation}
while the speed of sound is
\begin{equation}
c_s=\sqrt{\partial P\over \partial \rho}\sim\sqrt{k_B T\over
M}\sim v_T
\end{equation}
with $v_T$ the thermal velocity. These lead to the combination
\begin{equation}
L_\eta={\eta \over \rho c_s}\sim l_{MFP}
\end{equation}
Unlike the mean-free-path which is conceptually intuitive but
practically not easily computable or measurable (e.g. for fluids),
the length $L_\eta$ is well-defined by macroscopic properties of
the fluid and thus of practical use.

We emphasize that involving the speed of sound in the definition
of length scale $L_\eta$ is essential. This can be seen already
from a dimensional argument: $\eta/\rho$ or $\eta/(w/c^2)$ has the
dimension of $\rm[Length]^2\,[Time]^{-1}$. To turn this into a
length scale one needs to divide it by a quantity of the dimension
$\rm[Length]\,[Time]^{-1}$ i.e. a velocity. The natural choice
here is the speed of sound as a characteristic of the macroscopic
matter. The speed of light, $c$, on the other hand is \emph{not}
suitable here since it is neither a property of any specific
substance nor should it be of any relevance to non-relativistic
fluids, such as water. Since the speed of sound changes
considerably close to a phase-transition or a rapid crossover, its
inclusion in the fluidity measure $\mathcal{F}$ and in the
effective mean free path $L_\eta$ is essential.

Next we need to introduce another meaningful length scale to make
a dimensionless ratio: this becomes a necessity when comparing
fluids at vastly different scales. In case of the well known
dimensionless ratios like the Knudsen and Reynolds numbers, an
{\em external} length scale characteristic of the fluid motion is
introduced, like the diameter of a pipe or the size of a moving
object inside the fluid, etc. In the context of relativistic
fluids created in heavy ion collisions, there have been scaling
studies of collective flow for lower energy collisions based on
the Reynolds number in \cite{Reynolds} and more recently for
higher energy collisions based on the Knudsen number in
\cite{Knudsen} (see also related discussions in
\cite{Betz:2008me}). The external length scale in those numbers,
however, is not an {\em intrinsic} property of the fluid that we
would like to invoke for comparing fluids across vastly different
scales. Instead, a de-correlation length of certain
density-density spatial correlator gives a natural scale of
short-range order in the system. In most cases this de-correlation
length is simply set by the inter-(quasi-)particle distance. For a
non-relativistic fluid ``particles'' and their number density $n$
are well-defined, and thus the inter-particle distance is also a
well-defined length scale:
\begin{eqnarray}
L_n \equiv \frac{1}{n^{\frac{1}{3}}}
\end{eqnarray}
For relativistic fluid it is less straightforward. For example a
QGP with $\mu_B=0$ and thus $n_B=0$ can still have substantial
numbers of quarks and gluons. For such a relativistic fluid a simple
estimate can be made via the entropy density, i.e. $n\sim
\frac{s}{4k_B}$. Another way is to calculate the de-correlation
length of e.g. the correlator $<T_{00}(\vec x,0)T_{00}(\vec 0,0)>$
as has been done in recent lattice work \cite{Meyer:2008dt} which
finds a short range order about $0.6/T$ for QGP in $1-2Tc$, in
reasonable agreement with estimate from $n\sim \frac{s}{4k_B}$.
There could still be academic examples where the entropy density
can be infinite while the short-range order does {\em not} vanish:
AdS/CFT offers such an example in which the entropy density goes
as $s\propto N_c^2 T^3 \to \infty$ (in the large $N_c$ limit) implying
$1/n^{1/3} \to 0$ while spatial correlators like $<T_{\mu
\nu}(\vec x,0)T_{\mu \nu}(\vec 0,0)>$ give a non-zero
de-correlation length $\sim 1/T$. We will return to these issues
in the Subsection \ref{sec:sub_3_2}.

Finally by taking a ratio of the two length scales, we arrive at a
fluidity measure
\begin{eqnarray}
{\mathcal F}\equiv \frac{L_\eta}{L_n} \quad .
\end{eqnarray}
Below we will first show that the measure works well for
non-relativistic fluids and bears certain universality for
critical fluids. We will then show that the so-called
supercritical fluids have even better fluidity. At the end we
present a comparison of various interesting fluids. Following
\cite{Csernai:2006zz,Lacey:2006bc}, we have extensively used the
measured data for various fluids from the NIST WebBook
\cite{NIST_webbook}.

\subsection{Critical Fluids}

\begin{figure*}
\center{
    \hskip 0in\includegraphics[width=8cm]{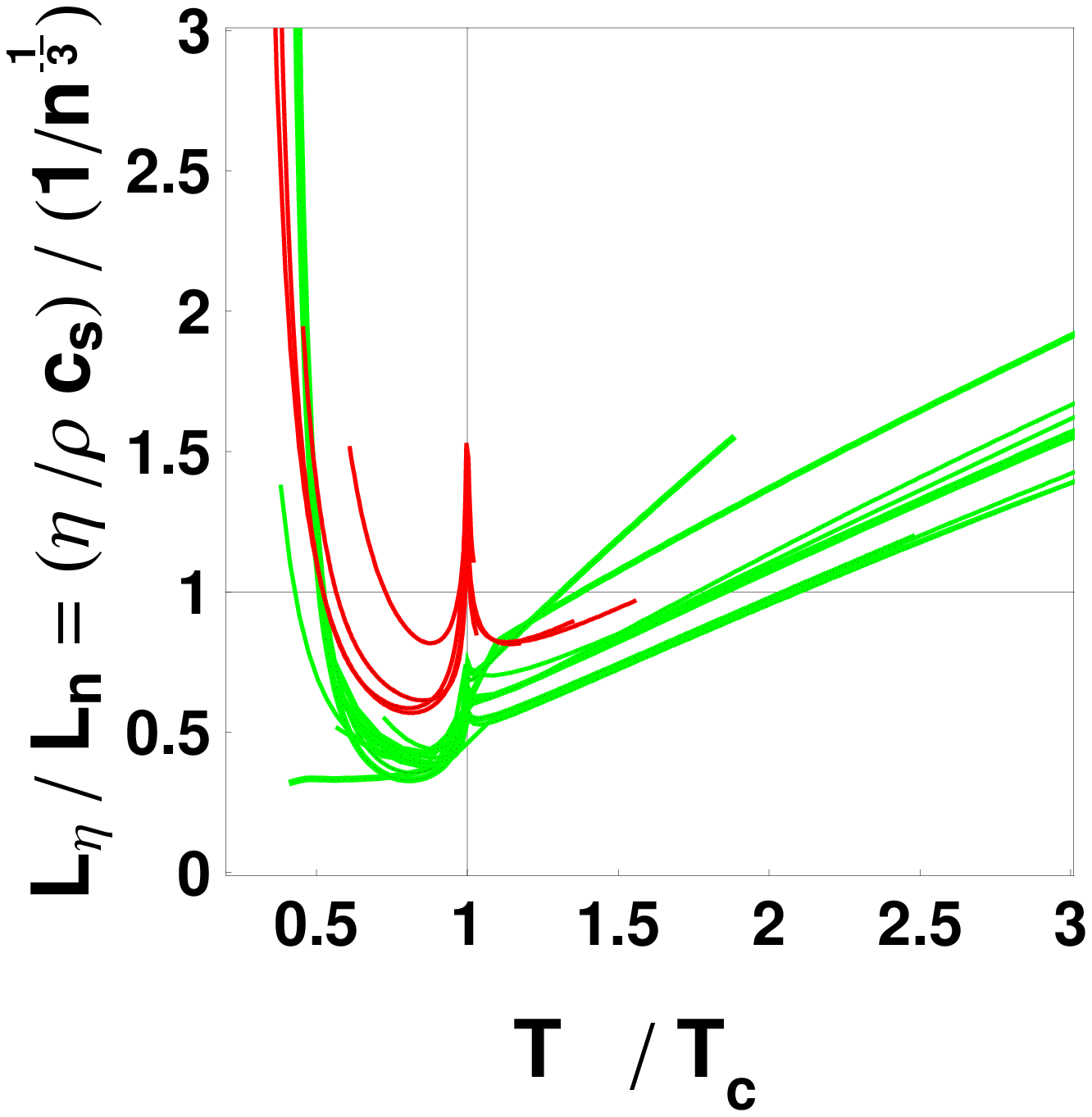}
\hskip 0.3in\includegraphics[width=8cm]{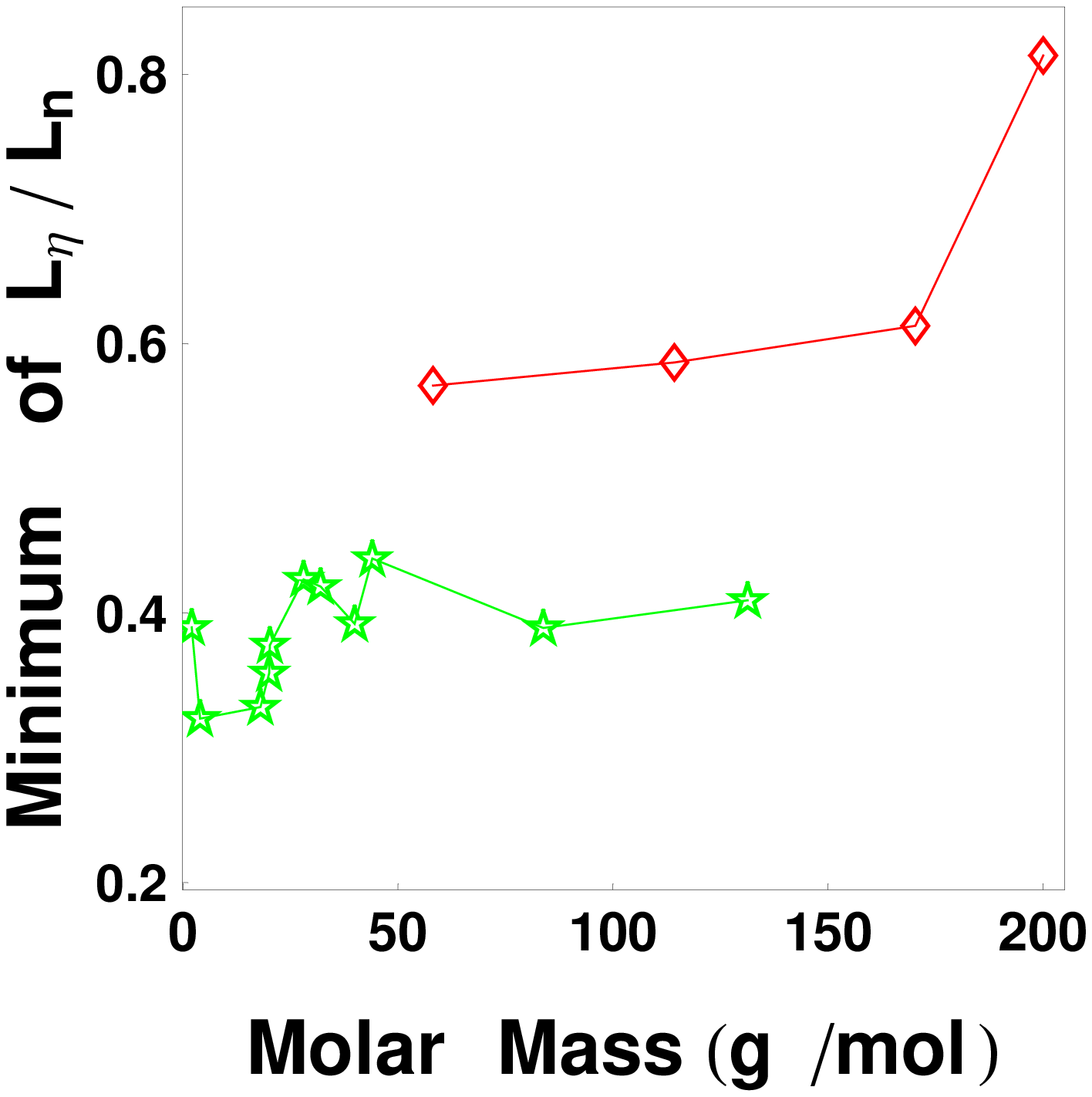}}
    \vskip 0.01in
 \caption{\label{fig_critical_fluids}
(Color online) (left panel) Fluidity measure ${\mathcal F}=
{L_\eta}/{L_n}$ versus $T/T_c$ for fifteen different substances at
fixed critical pressure $P=P_c$, see text for more details. The
sharp peaks centered at $T_c$ are due to the actual {\em
divergence} of the shear viscosity at the critical point. (right
panel) The minimum of the fluidity measure ${\mathcal F}$ for
various substances (as obtained from respective curves in the left
panel) versus their molar masses, with the stars (from left to
right) for  $H_2$, $^4 He$, $H_2 O$, $D_2 O$, $Ne$, $N_2$, $O_2$,
$Ar$, $CO_2$, $Kr$, $Xe$, and the diamonds(from left to right) for
$C_4 H_{10}$, $C_8 H_{18}$, $C_{12}H_{26}$, $C_4 F_8$. }
\end{figure*}

We first examine various fluids at fixed critical pressure
$P=P_c$. In Fig.\ref{fig_critical_fluids}(left), the proposed
fluidity measure ${\mathcal F}= {L_\eta}/{L_n}$ is plotted for {\em fifteen}
different substances at their respective critical pressure $P_c$,
including: Hydrogen ($H_2$), Helium-4 ($^4 He$), Water ($H_2 O$),
Deuterium oxide ($D_2 O$), Neon ($Ne$), Nitrogen ($N_2$), Oxygen
($O_2$), Argon ($Ar$), Carbon Dioxide ($CO_2$), Krypton ($Kr$),
Xenon ($Xe$), Isobutane ($C_4 H_{10}$), Octane ($C_8 H_{18}$),
 Dodecane ($C_{12}H_{26}$), Octafluorocyclobutane ($C_4 F_8$).
These substances cover a wide range of molar mass, chemical
structure and complexity, with their respective critical
temperature $T_c$ and pressure $P_c$ differing by orders of
magnitude. Despite such huge differences, their fluidity curves
resemble each other not only in shape but even {\em
quantitatively}. In particular in their good liquid regime ---
roughly the ``valley'' region at $\sim 0.7-1\, Tc$ --- they all
show amazingly similar fluidity. To further expose the similarity,
 in the right panel of  Fig.\ref{fig_critical_fluids} we show
the value
of the
fluidity measure ${\mathcal F}$ at its minimum versus the substances' molecular
molar masses. From this plot,
roughly two bands can be identified: the green stars spread in a
narrow band of ${\mathcal F}\in (0.3,0.45)$ while spanning
two-orders-of magnitude in molar mass which include all 11
non-organic substances; the red diamonds with roughly twice bigger
${\mathcal F}$, include 4 organic substances with much more
complicated molecular structures (e.g. chains) which, not
surprisingly, lead to more dissipation. Even so, the splitting in
fluidity between the two bands is merely a $\hat O(1)$ factor
rather than any order-of-magnitude difference. Nevertheless as one
can imagine, with increasing chemical complexity and molecular
mass, the fluidity of more complex systems like e.g. engine oil
may deviate significantly from what are shown in the figure.

To conclude the study of critical fluids, there appears to be
certain universality of the newly proposed fluidity measure
$\mathcal F$, indicating that ``a good fluid is a good fluid''
despite many other details regarding the microscopic degrees of
freedom.

\begin{figure*}
\begin{center}
    \hskip 0in\includegraphics[width=8.cm]{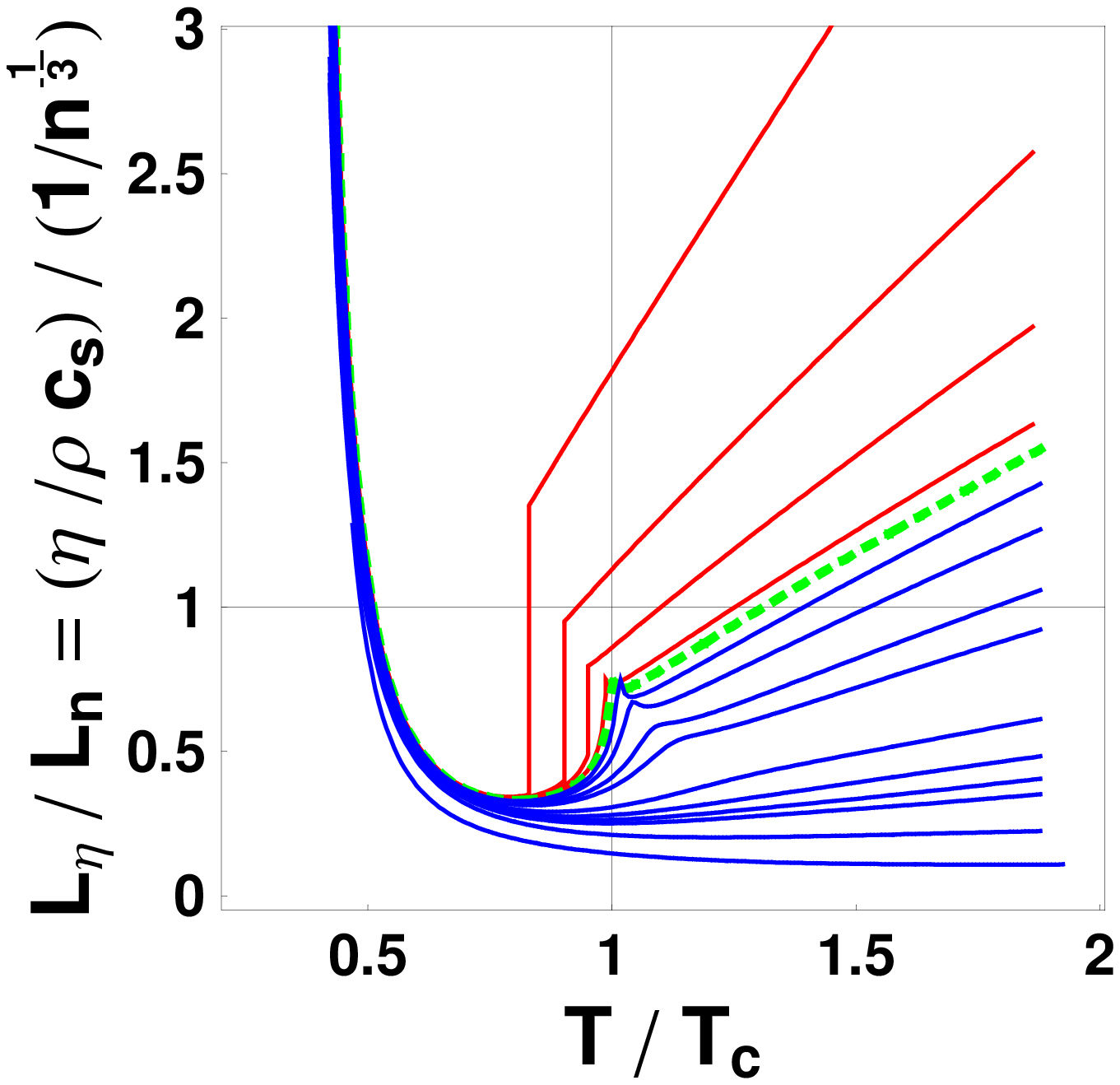}
 \hskip 0.3in    \includegraphics[width=8.cm]{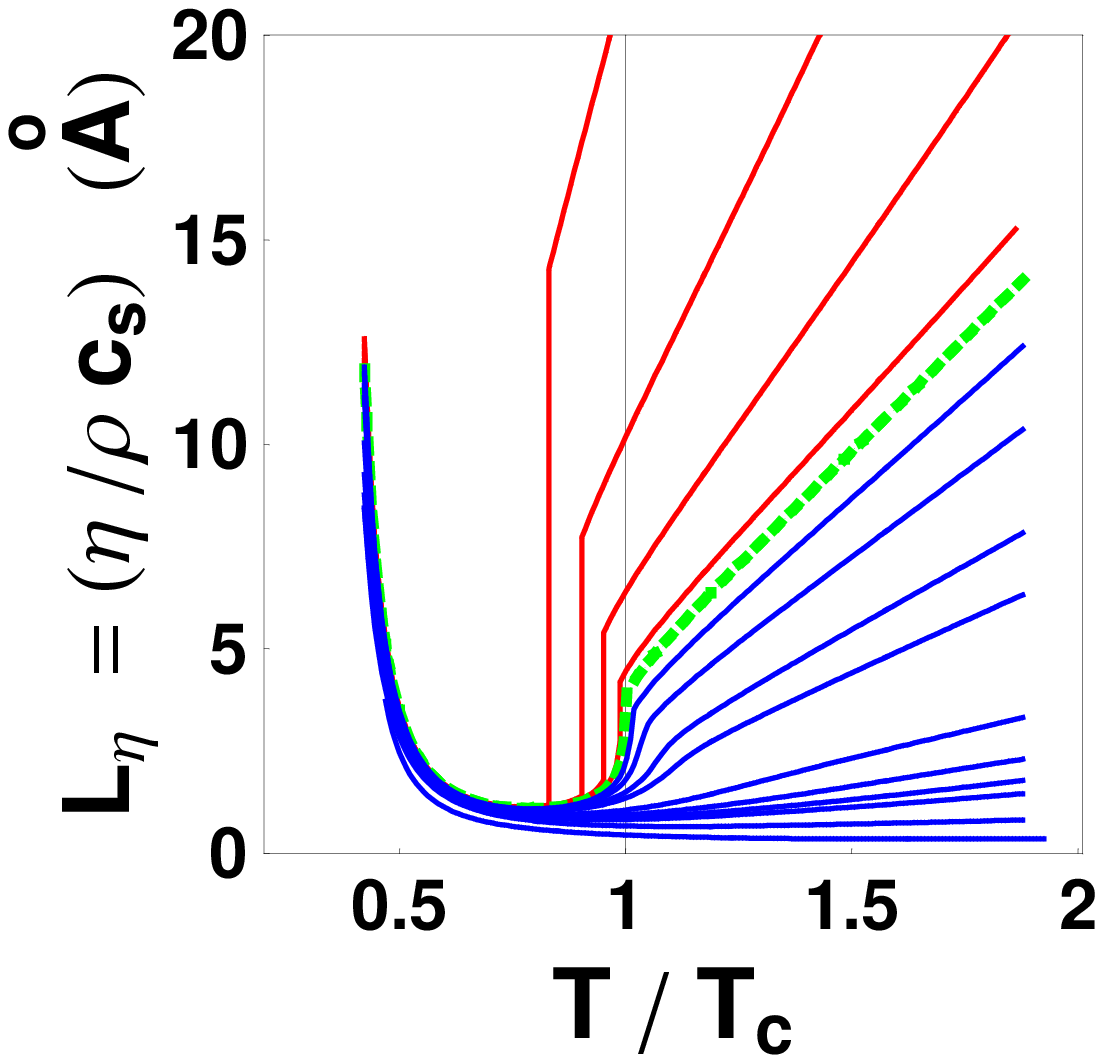}
\end{center}
 \caption{
\label{fig_water_run_p}
 (Color online) The dimensionless fluidity measure $\mathcal F$ (left panel) and
  the length scale, $L_\eta$
(in units of \AA ) (right panel) versus $T/T_c$
 for  water at fifteen different fixed pressure  values. In both panels,
 the dashed (green) curve is for
 fixed critical pressure $P=P_c=22\,\rm MPa$, while the four
  solid (red) curves above it are for $P<P_c$, and the ten solid (blue) curves below it
  are for $P>P_c$, with the respective pressure values for each curve (from top down)
  being $P=5,10,15,20,22,25,30,40,50,100,150,200,250,500,1000\, \rm{MPa}$.}
\end{figure*}

\begin{figure}
\begin{center}
\includegraphics[width=8.cm]{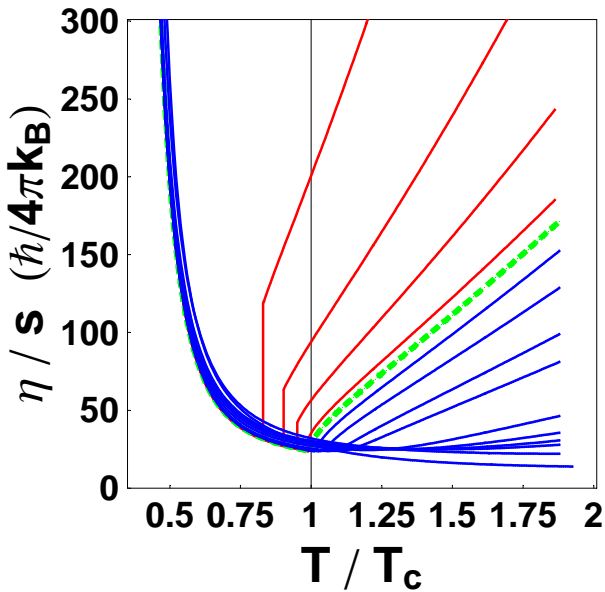}
\end{center}
    \vskip 0.01in
 \caption{\label{fig_water_eta_s}
(Color online) The shear-viscosity-entropy-density-ratio $\eta/s$
(in unit $\hbar/(4\pi k_B)$)  versus $T/Tc$ for water at fifteen
different fixed pressure values. The dashed (green) curve is for
 fixed critical pressure $P=P_c=22\,\rm MPa$, while the four
  solid (red) curves above it are for $P<P_c$, and the ten solid (blue) curves below it
  are for $P>P_c$, with the respective pressure values for each curve (from top down)
  being $P=5,10,15,20,22,25,30,40,50,100,150,200,250,500,1000\, \rm{MPa}$.}
\end{figure}

\subsection{Supercritical Fluids}

While much attention has been paid to critical fluids, the
fluidity of a fluid with significantly larger pressure $P>>P_c$
has been little discussed in the context of heavy ion collisions
and QCD matter: so let us next explore this region. In the
literature, this region is often referred to as the
``supercritical fluid'' region \cite{supercritical}, defined on a
typical substance's $T-P$ phase diagram (with critical point
$T_c,P_c$) as:
\begin{eqnarray}\label{eqn_super_critical}
supercritical:\quad T>T_c \,\, \& \,\, P>P_c
\end{eqnarray}
In particular we want to explore the deeply supercritical region
with $P>>P_c$. Taking water as an example, in
Fig.\ref{fig_water_run_p}(left) we plot the fluidity measure
$\mathcal F$ versus $T/Tc$ for fifteen different fixed pressure
values ranging from $5\,\rm MPa$ all the way to $1000\,\rm MPa$
(note for water $P_c=22\,\rm MPa$). The dashed (green) curve is
for fixed critical pressure $P=P_c=22\,\rm MPa$, while the four
  solid (red) curves above it are for $P<P_c$, and the ten solid (blue) curves below it
  are for $P>P_c$. As can be
 seen from the plot, the curves change shape gradually and
 the fluidity becomes better and better with increasing $P$. The
 ``valley'' where $\mathcal F$ remains small and relatively flat
 becomes much wider and eventually flattens out
 substantially above $T_c$ for $P>>Pc$. To quantify the
remarkable
 fluidity of the supercritical fluid, we
 note that the minimum on the $P=P_c$ curve has a value for the fluidity
${\mathcal
 F}_{min}(Pc)\approx 0.33$ while the minimum for the  $P=1000\, {\rm MPa}\approx
 45P_c$ curve is at ${\mathcal F}_{min}(45Pc)\approx 0.11$, getting
 smaller by a factor of 3 and remaining close to the minimum within
 a rather broad temperature region!

It is also interesting to examine whether the length scale
$L_\eta$ itself shows similar trends, as $L_\eta$ is the essential
scale for discussing the applicability of hydrodynamics where it
is to be compared with the external scale characterizing the
variation of flow field. In the right panel  of
Fig.\ref{fig_water_run_p} we plot $L_\eta$ itself for the same
conditions and find that, similar to
the fluidity measure $\mathcal F$, the scale $L_\eta$ also becomes
considerably smaller in  supercritical water.

It should be mentioned that the same observation is also true for
other fluids that we examined, like Helium-4, Nitrogen, etc.
Furthermore such behavior is not specific to our fluidity measure. In
Fig.\ref{fig_water_eta_s} we plot the widely used ratio of shear viscosity
over entropy-density, $\eta/s$, which show the same qualitative behavior.

The main lessons, as we emphasize again, are (a) for a given
substance, the best fluidity is {\em not} necessarily achieved
close to the critical point and (b) when going deeper into the
supercritical regime its fluidity becomes much better.

\subsection{Comparison of Various Fluids}
\label{sec:sub_3_2}

\begin{figure}
\begin{center}
    \hskip 0in\includegraphics[width=8.cm]{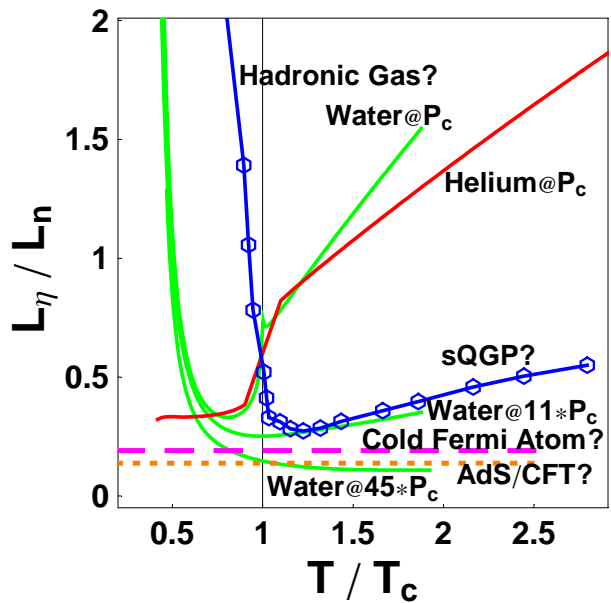}
    \vskip 0.01in
    \end{center}
 \caption{\label{fig_comparison}
(Color online) Comparison of fluidity measure for various fluids
(see text for more details). The curves with question marks
indicate current estimates of the respective fluidity with
possible uncertainty, while the curves for Helium at $P_c$ and for
water at $P_c,11P_c,45P_c$ are from actual data.}
\end{figure}

Finally we attempt to compare various fluids of great current
interests in terms of the new fluidity measure $\mathcal F$ we
have proposed and studied above. The results are shown in
Fig.\ref{fig_comparison}. Below we give the details of how the
curves for QCD, Cold Fermi Atom, and AdS/CFT are obtained.

For the QCD system, we have used a parametrization of the
viscosity $\eta$ by Hirano and Gyulassy in \cite{Hirano:2005wx}:
one uses $\eta/T_c^3 \approx T/T_c$ for hadronic gas (H.G.) below
$T_c$, and $\eta/T_c^3\approx (T/T_c)^3
[1+\mathcal{W}(T)\mathbf{ln}(T/T_c)]^2$ for the quark-gluon
plasma(sQGP) in the temperature interval  $T/T_c \in [1-3]$ with\\
$[\mathcal{W}(T)/(4\pi)]=(9\beta_0^2)\cdot[80\pi^2 K_{SB}
\mathbf{ln}(4\pi/g^2(T))]^{-1}\,\,$ interpolating to pQCD results
for $T>>T_c$, where the parameters are given as $\beta_0=10$, $K_{SB}=12$ and the running coupling\\
$[g^2(T)]^{-1}=(9/8\pi^2)\mathbf{ln}(2\pi T/\Lambda)+(4/9\pi^2)
\mathbf{ln}(2\mathbf{ln}(2\pi T/\Lambda))\,\,$ (with
$\Lambda\approx 190MeV$) ( see
\cite{Hirano:2005wx}\cite{Csernai:2006zz} for more details). The
enthalpy density $w=\epsilon+p$ and speed of sound $c_s$ are taken
from recent lattice results by Karsch et al \cite{Cheng:2007jq}
for 2+1 flavor QCD with $m_\pi\approx 220MeV$. As we mentioned
before, $L_n$ is estimated by $1/(s/4k_B)^{1/3}$ with the entropy
density also taken from \cite{Cheng:2007jq}.

For the strongly coupled AdS/CFT system, the shear viscosity is
well known to be $\eta/s=1/(4\pi)$ \cite{Policastro:2001yc}. As we
also pointed out before, there is a short-range order at the
length scale $L_n\sim 1/T$ however the pre-factor is not
accurately determined. We simply use $L_n=1/T$ as an estimate.
This gives the fluidity ${\mathcal F}=\sqrt{3}/(4\pi)\approx
0.138$.

For the Cold Fermi Atom gas, its shear viscosity has been measured
near its Feshbach resonance by Thomas et al in
\cite{fermi_atom_viscosity} for a certain range of system energy
by analyzing the damping of collective modes in the atomic cloud
(see also related work in \cite{Schaefer}). As a benchmark, we
take the lowest viscosity found from the measurement (see Fig.4 in
\cite{fermi_atom_viscosity}) which is $\eta\approx 0.214\, \hbar\,
n$. The speed of sound has also been measured by the same group in
\cite{fermi_atom_sound}, from which we take the value $c_s\approx
0.3632\, v_F$ near the Feshbach resonance. Since the Fermi velocity
$v_F=\hbar k_F/m$, the mass density $\rho=m n$, and the Fermi
momentum $k_F/n^{1/3}=(3\pi^2)^{1/3}$, we obtain for the fluidity
measure ${\mathcal
F}\approx 0.214/(0.3632\cdot(3\pi^2)^{1/3})\approx 0.191$.

The question marks in Fig.\ref{fig_comparison} indicate that the
current estimates presented above may carry sizable uncertainties,
and we expect the knowledge on these systems will become more
accurate with time.

We finally come to the comparison in Fig.\ref{fig_comparison}. As
one can see,  the critical fluids (water and Helium-4 at $P_c$)
have a somewhat worse fluidity than the QCD, AdS/CFT and Cold
Fermi Atom systems. Supercritical water at $P=11Pc$, on the other
hand,  already has a fluidity comparable to the QGP while at
$P=45Pc$ the fluidity of supercritical water appears to be better
than that of the recently discussed  ``nearly perfect fluids''.

\section{A Supercritical QCD Fluid at RHIC?}
\label{sec:section_4}

The fluidity study in the previous section, in particular on the
supercritical fluid, naturally leads to the following interesting
question: {\em Are we observing a supercritical QCD fluid at
RHIC?}

\begin{figure}
\begin{center}
 \includegraphics[width=8.5cm]{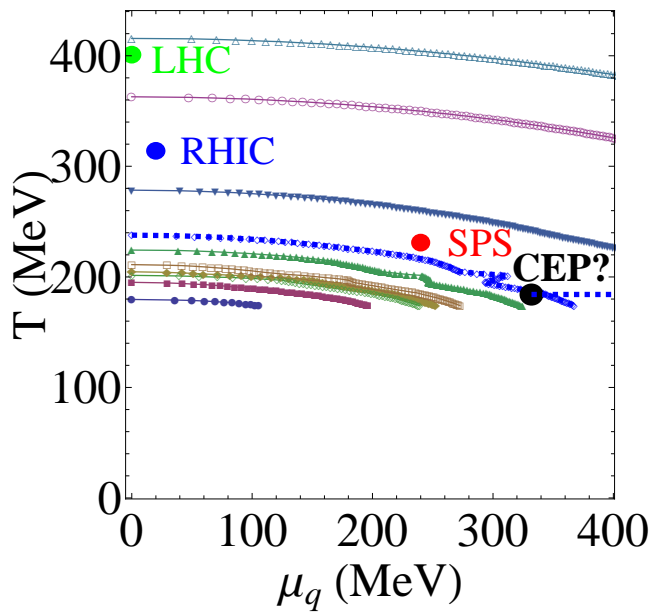}
\end{center}
    \vskip 0.01in
 \caption{\label{fig_QCD}
(Color online) Schematic isobaric contours (i.e. with constant
pressure along each line) on the QCD $T-\mu$ phase diagram with
the filled black circle indicating a possible position of the {\em
hypothetical} Critical-End-Point (CEP); the dashed blue horizontal
short line stretching from the CEP to the right is included to
indicate the $T=T_{CEP}$ boundary (see text for more details). The
filled red, blue and green circles indicate estimates of the
initial $(T,\mu)$ reachable at SPS, RHIC and LHC, respectively.
 }
\end{figure}

As is well known, what is usual referred to as $T_c\approx 170\sim
190\rm MeV$ in QCD is {\em not} a true second-order phase
transition temperature but rather the temperature for a rapid
crossover at $\mu=0$ \cite{Cheng:2007jq}. There is, however, a
(hypothetical) Critical-End-Point(CEP) at $(T_{CEP},\mu_{CEP})$ on
the QCD phase diagram marking the end of a plausible first-order
phase transition at low $T$ but high $\mu$ (see e.g.
\cite{Stephanov:2007fk} and references therein). Accurate
determination of the CEP from lattice QCD \cite{Fodor:2001pe,de
Forcrand:2003hx,Gavai:2004sd,Allton:2005gk,Philipsen:2005mj} and
unambiguous observation of the CEP from heavy ion collisions
\cite{Lacey:2006bc,Stephanov:1998dy,Lacey:2007na} are
among the most interesting and exciting goals of QCD research.
While currently the position of the CEP is not well constrained,
it is expected to have lower temperature than that of the QCD
crossover transition at vanishing baryon density, $T_{CEP}<T_c$.
Being very aware on the present uncertainty of the actual location
of the QCD critical end point, let us, solely for demonstration
purposes and for the sake of the argument, assume that its
location is close to the estimate of ref. \cite{Gavai:2004sd},
which suggests that $T_{CEP}\approx 0.94 T_c,\mu_{CEP}\approx 1.8
T_{CEP}$. This location is indicated  in Fig.\ref{fig_QCD} as the
filled black circle along with a question mark.

Next if we adopt the definition of supercriticality as in
Eq.(\ref{eqn_super_critical}) for QCD, we need to know the
constant-pressure lines on the QCD phase diagram typically plotted
in terms of $T-\mu$. To schematically construct these lines, we
make use of the Taylor expansion of the pressure $P(T,\mu)$ with
respect to $\mu$:
\begin{eqnarray}
P(T,\mu) = T^4 \times && {\bigg [}  P_0(T) + \frac{1}{2}
\chi_2^B(T)\, \left(\frac{\mu}{T}\right)^2  \nonumber \\
&& \,\,\, + \frac{1}{24} \chi_4^B(T)\,
\left(\frac{\mu}{T}\right)^4 + ... {\bigg ]}
\end{eqnarray}
with $\chi_{2n}^B \equiv \frac{\partial^{2n}(P/T^4)}{\partial
(\mu/T)^{2n}} {\big |}_{\mu=0} $ the baryonic susceptibilities
\cite{Gavai:2004sd,Allton:2005gk,Cheng:2008zh,Liao:2005pa,Koch:2008ia}.
In order to construct the equal-pressure
lines shown in Fig.\ref{fig_QCD} we have used the lattice QCD
results from Cheng et al
for $P_0(T)$ \cite{Cheng:2007jq}  and   $\chi_2^B,\chi_4^B$
\cite{Cheng:2008zh}.

The blue curve in Fig.\ref{fig_QCD} going right through the
indicated CEP point indicates the constant-pressure line with
fixed critical pressure $P=P(CEP)$ which, together with the
$T=T_{CEP}$ line (dashed blue horizontal one stretching from the
CEP to the right) form the boundary above which there is the {\em
QCD supercritical region}. We emphasize that above the boundary
the pressure $P\sim T^4$ increases very rapidly. As a result the
QGP quickly enters deeply into the supercritical regime where
$P\gg P(CEP)$. To give an idea: the pressures of the lines in
units of the critical pressure $P(CEP)$ (value on the blue curve)
are (from bottom to top) $0.05$, $0.08$, $0.13$, $0.23$, $0.27$,
$0.35$, $0.63$, $1$, $2.7$, $10$, $18$, respectively. Of course
with improved lattice results for pressure, CEP and (higher order)
susceptibilities a more accurate equal-pressure map for QCD matter
will become available.

We now discuss a few implications for heavy ion collisions
experiments. As a precaution, the following discussions are by no
means intended to provide quantitative statements but rather
qualitative yet interesting ideas.
\\
(I) At current RHIC energy ($\sqrt{s}=200\,\rm AGeV$),
Fig.\ref{fig_QCD} indicates that most of the QGP phase of the
created matter is likely in the supercritical region. This may
very well be the reason why such good fluidity has been observed.
If  a relationship between  good fluidity and  supercriticality is
indeed true for the QCD matter just as it is for  water, then
there will be {\em a sensitive dependence of the fluidity on  the
actual position of the CEP.} For example, the matter created at
RHIC might  have most of its evolution being in the supercritical
region in case $T_{CEP}$ and $P_{CEP}$ are considerably lower than
indicated in the Fig.\ref{fig_QCD}. On the other hand, the matter
created at SPS ($\sqrt{s}\simeq 20 \,\rm GeV$) seems to have its
initial pressure and temperature already rather close to the
critical one which would result in poor fluidity. In principle
these regions (supercritical, near-CEP, below-CEP) of very
different fluidity can be accessed experimentally by tuning the
center of mass energy $\sqrt{s}$ which changes the initial
densities (see e.g. \cite{Kestin:2008bh}) as well as  the
freeze-out points (see e.g.
\cite{BraunMunzinger:2001ip}\cite{Randrup:2009gp}) in such
collisions. To which extend this difference in fluidity transforms
into measurable effects, such as non-hydrodynamic behavior and
 thus less elliptic flow, is a non-trivial question. As we will discuss in the
next section, besides the fluidity the actual size of the system
is essential for the applicability of hydrodynamics. Our estimates
(see Fig.\ref{fig_QCD_Leta}) show that  even for the very good
fluidity assumed for a strongly interacting QGP, denoted as
``sQGP'' in Fig.\ref{fig_comparison}, a hydrodynamic description
for the fireball expansion is a best marginal at SPS energies.
The reason is that, according to our estimate, the system starts
close to the phase transition and, thus, reaches the hadronic
phase while still comparatively small in size. Consequently, it
may be difficult to tell if the system enters a regime of reduced
fluidity from flow studies alone. Of course if our estimates are
wrong and one finds evidence that the initial entropy density
reached is considerably larger, a change in the fluidity may show
up in the systematics of elliptic flow measurements, such as its
dependence on beam energy and system size.
\\
(II) Based on the possibility of a mostly supercritical QCD matter
to be created in experiments at the Large Hadron Collider (LHC)
(see Fig.\ref{fig_QCD}), one is led  to predict {\em an even
better fluidity to be observed at LHC than at RHIC}. And since for
the same fluidity the conditions for hydrodynamics are more
favorable for LHC energies than for RHIC, we would expect nearly
ideal hydrodynamic evolution at LHC.

\section{Applicability of Hydrodynamics in Heavy Ion Collisions}
\label{sec:section_5}

In this section we turn to a discussion of applicability of
hydrodynamics in heavy ion collisions, particularly for SPS, RHIC
and LHC. As already discussed in the introduction, the fluidity of
the underlying matter is {\em not} the only determining factor:
the ``best'' (in terms of fluidity) fluid may cease to flow if put
into  a sufficiently narrow pipe. In other words, the
applicability of hydrodynamics depends equally on both the
\emph{internal} properties of the substance, i.e. its fluidity
$\mathcal{F}$ or $L_\eta$, and the \emph{external} settings, i.e
the typical length scale $L_s$ of the variation of the flow field
in the system under consideration \footnote{We re-emphasize that
the length scale $L_\eta$ is a well defined quantity depending
only on macroscopic variables that are measurable and/or
calculable. Furthermore unlike the mean-free-path which depends on
quasi-particle picture (see e.g. discussions in
\cite{Betz:2008me}), the length scale $L_\eta$ remains a useful
and meaningful scale even for strongly coupled systems.} : one
useful criteria, equivalent to the famous Knudsen number, is a
ratio of the two $L_\eta / L_s$.

In order to derive a quantitative criterion, we again use the
example of sound dispersion discussed before. By requiring that a
propagating sound wave has to complete at least one full period
before its amplitude damps by a factor $e^{-1}$ due to the
imaginary part, we arrive at a {\em minimal} wave length
$\lambda_{min}=\frac{8\pi^2}{3} L_\eta$. For a sound wave the
length scale characterizing the flow field variations  simply the wavelength,
$L_s \approx \lambda$. Thus we arrive at the following criterion  for the
applicability of hydrodynamics:
\begin{equation} \label{eqn_criteria}%
\frac{L_\eta}{L_s} < \frac{3}{8\pi^2} \approx 0.038
\end{equation}%

\begin{figure*}
 \includegraphics[width=7.5cm]{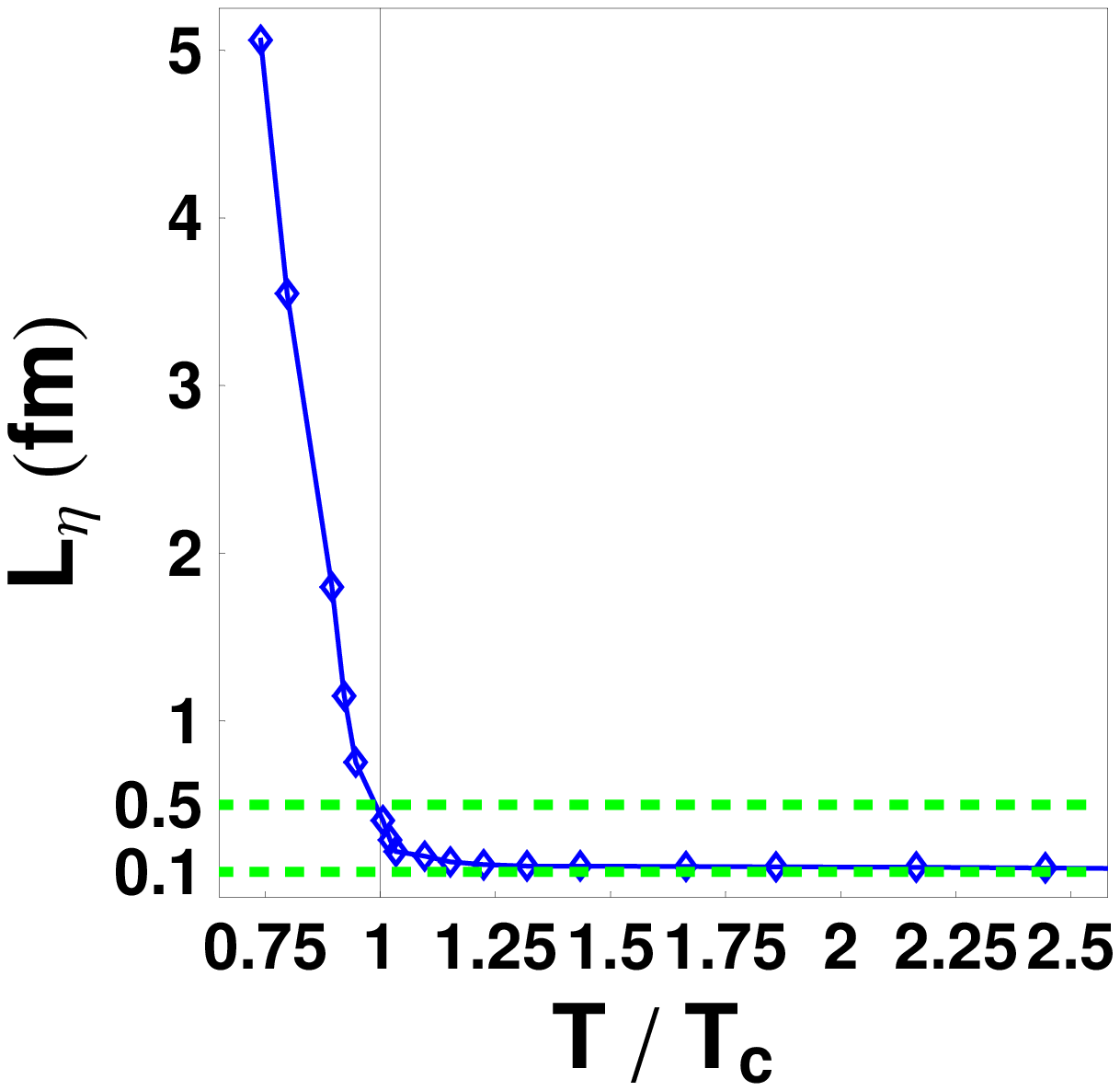}
\hskip 0.3in \includegraphics[width=8.cm]{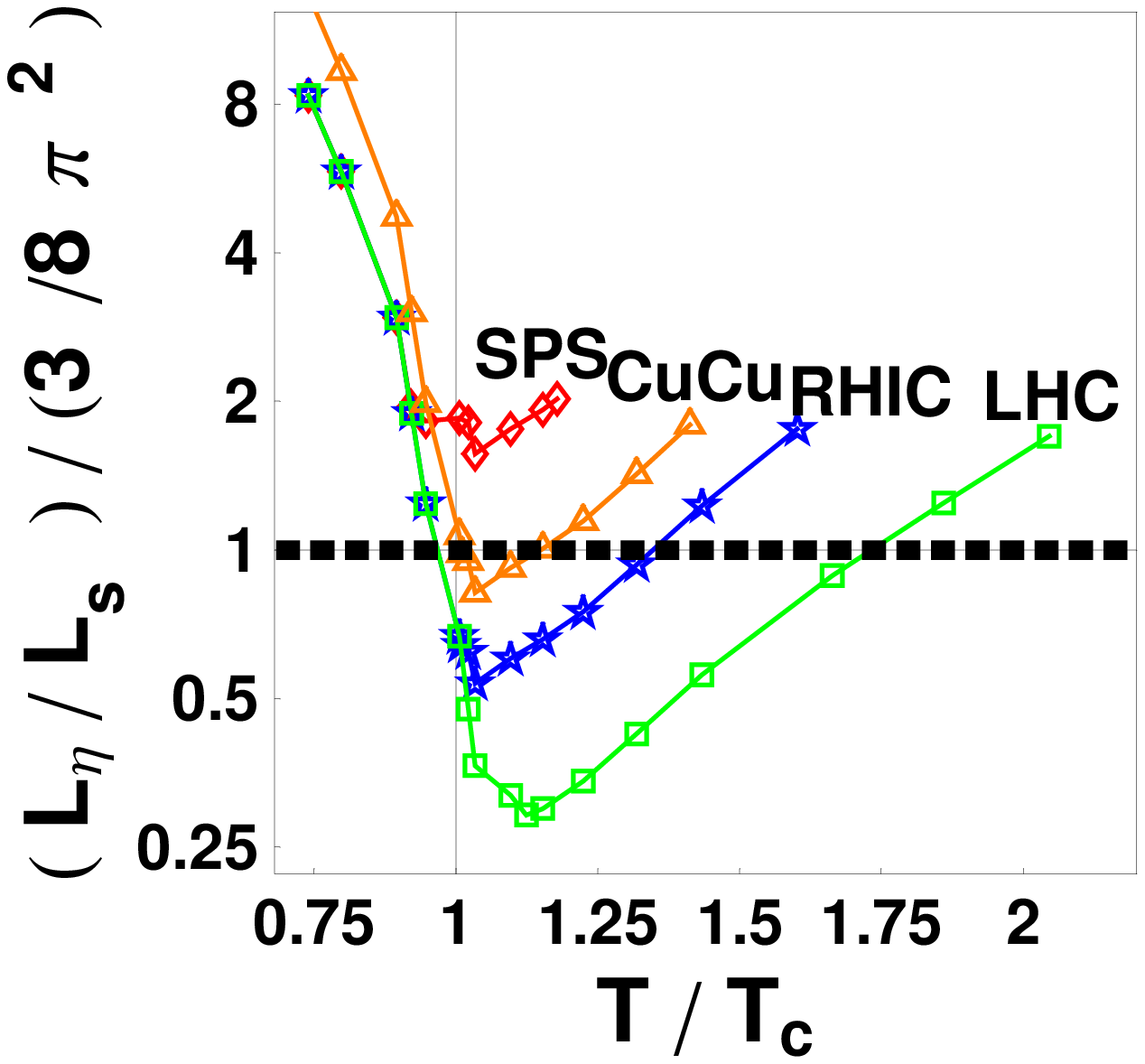}
    \vskip 0.01in
 \caption{\label{fig_QCD_Leta}
(Color online) (left panel) The length scale
$L_\eta=\frac{\eta}{(w/c^2) c_s}$ (in unit $\rm fm$) versus
$T/T_c$ for QCD at $\mu=0$, with $\eta$ from parametrization in
\protect\cite{Hirano:2005wx} and $w,c_s$ from lattice data
\protect\cite{Cheng:2007jq}. The dashed horizontal lines at
$0.1\,\rm fm$ and $0.5\,\rm fm$ are included to guide the eyes.
(right panel) The criteria (\protect\ref{eqn_criteria}) for
applicability of hydrodynamics versus $T/T_c$ at SPS(red
diamonds), RHIC(blue stars), and LHC(green boxes), with the thick
dashed horizontal line indicating the borderline below which
hydrodynamics may be a good description according the criteria
(see text for more details). A line for CuCu(orange triangles)
collisions at RHIC energy ($200\, \rm AGeV$) is also presented. }
\end{figure*}

The length $L_\eta(T)$ for QCD (at $\mu=0$), shown in
Fig.\ref{fig_QCD_Leta}(left), is determined as described section
\ref{sec:sub_3_2}. It relies on a parametrization for
$\eta(T)$ from the Hirano-Gyulassy \cite{Hirano:2005wx}, and we note that
the actual shear viscosity of QCD matter, once it is known,  may be quite
different.
The
temperature dependence of $L_\eta$ shown in the plot features a
steep drop from the hadronic phase to about $T\simeq 1.1 T_c$,
followed by a rather constant value for temperatures above,
$T\gtrsim 1.1 T_c$. One may expect $L_\eta$ becomes large again at
very high temperatures $T>>T_c$ where a weakly coupled QGP
describable by pQCD takes over. In principle $L_\eta$ also depends
on the baryon-number chemical potential, $\mu_B$, and this
dependence shall be taken into account when discussing collisions
at low energy (such as SPS) where $\mu_B$ could be sizable.
However currently very little is known about such dependence, and
we here use the zero $\mu$ result also for SPS.

We next give a rough estimate for the typical length scale $L_s$
of flow field variance for heavy ion collisions at different
$\sqrt{s}$ \footnote{In this discussion we focus only on central
collisions which make the estimation easier. Generally speaking
when going from central to peripheral collisions one expects
hydrodynamics to be less and less applicable.}. At early times,
right after the collision, the system size is limited mainly by
its longitudinal extent and the scale governing the variation
in the fluid can be estimated to be about $2\tau$. At late time
when $\tau$ becomes larger than the nuclear radius $\sim R_A$ the
limiting scale is set by the transverse size of the fireball which
is slightly larger than $R_A$ and slowly grows. For the purpose of
our rough estimate, we adopt a simple prescription:
 $L_s \approx 2\tau$ for $\tau<R$ and $L_s \approx 2R$ for $\tau\ge
R$ with $R=8\, \rm fm$ for AuAu and PbPb collisions.
For the collision dynamics
which
determines the temperature evolution $s(\tau)$ (or inversely
$\tau(s)$), we simply use the Bjorken flow relation
$s(\tau)=s_0(\tau_0)\cdot \tau_0 / \tau$, which up to $\tau \sim
R$ is a reasonable approximation. For the initial
condition we follow estimates used in hydrodynamic modelling
(see e.g. \cite{Kestin:2008bh}) and assume equilibration at
$\tau_0=1\, \rm fm$, with
fireball center entropy densities (in central collisions) to be %
$s_0=24\, \rm fm^{-3}$ at SPS, $s_0=70\, \rm fm^{-3}$ at RHIC, and
$s_0=154\, \rm fm^{-3}$ at LHC.

Finally we turn to the discussion of the applicability of
hydrodynamics given the criteria of Eq.\ref{eqn_criteria}. The
ratio $L_\eta / L_s$ in units of $3/(8\pi^2)$ is plotted in
Fig.\ref{fig_QCD_Leta}(right) for SPS(red diamonds), RHIC(blue
stars), and LHC(green boxes). The thick dashed horizontal line
indicates the equality of the condition expressed by
Eq.\ref{eqn_criteria}: If the system is considerably below this
line a hydrodynamic description should be a reasonable
approximation. If it finds itself considerably above corrections
to hydrodynamics from higher orders in derivatives or an even
non-hydrodynamic description are called for. To quantify the
previous statement, if the system finds itself at a value of
$(L_\eta / L_s)/(3/8\pi^2) = 2$, such as the points for SPS
energies on Fig.\ref{fig_QCD_Leta}(left), then the amplitude of a
sound mode after propagating the distance of one wavelength will
be reduced
 by a factor of $e^2\simeq 7$, and for a value of  $(L_\eta / L_s)/(3/8\pi^2) =
5$ the reduction would be $e^5\simeq 150$!
 Given our estimate and the assumptions  which it is
based on, a hydrodynamic description is not a good approximation
for SPS, whereas for RHIC and even more so for LHC energies, it
seems to be more or less justified\footnote{We  note that even for
RHIC and LHC the first fm or so of the evolution is above the
``criteria line'' due to the rather small system size. Whether
this has a profound effect (such as in entropy generation) or not
is unclear at this time (see a very interesting discussion in
\cite{Lublinsky:2007mm}).}. Such differences from SPS to RHIC and
LHC lie in the different time evolution due to the different
initial densities: when cooling down into the region where
$L_\eta$ starts rising abruptly near and below $T_c$, the SPS
fireball still has a rather small size (longitudinally) while the
RHIC and LHC fireballs already becomes large with a size about
$2R$ and hence have their hydrodynamic evolution extended much
longer. While our estimates contain uncertainties and should by no
means  considered to be precise, we have provided a
semi-quantitative picture for evaluating the applicability of
hydrodynamics with varying $\sqrt{s}$, which implies at LHC the
fireball expansion shall be even better described by hydrodynamics
as compared to RHIC.

Besides the beam energy $\sqrt{s}$, a change of the system size
should also affect the conditions for the applicability of
hydrodynamics. For example, RHIC has done experiments with both
AuAu collisions and CuCu collisions, both at full energy, and it
would be interesting to examine and compare the applicability of
hydrodynamics for these two different systems. In
Fig.\ref{fig_QCD_Leta} (right) we have included a calculation of
the same applicability measure for CuCu (orange triangle)
collisions at $200\, \rm AGeV$, with a initial center entropy
density $s_0=45\, \rm fm^{-3}$ and a transverse size parameter
$R=5\, \rm fm$. The curve indicates that hydrodynamics is
marginally applicable for such collisions, as both the initial
density and the transverse size of the matter formed in CuCu
collisions are smaller that those in AuAu collisions at the same
energy.

\section{Summary}

In summary, we have discussed the fluidity of the hot and dense
QCD matter as produced in ultrarelativistic heavy ion collisions
in comparison with various other, well known fluids. In particular
we have suggested its possible supercriticality based on insights
gained from studying more conventional fluids like water. We have
discussed several aspects relevant for a proper comparison of
non-relativistic and relativistic  fluids, both  from
thermodynamics and hydrodynamics perspectives. A new fluidity
measure ${\mathcal F}=L_\eta/L_n$ is then proposed, which shows
certain universality in the good fluid regime for a remarkable
diverse set of critical fluids. We have further demonstrated that
the fluidity is enhanced in the supercritical fluid regime on a
fluid's phase diagram. These studies inspired us to conjecture
that the seemingly good fluidity of the QCD matter at RHIC may
actually be related to its supercriticality with respect to the
Critical-End-Point on the QCD phase diagram. This observation, if
true, has far-reaching consequences for heavy ion collisions
experiments: (a) the loss of such good fluidity at certain lower
beam energy which is sensitively related to the position of the
long sought CEP; (b) an even better fluidity may be expected at
higher beam energy, which will soon be tested by the LHC heavy ion
program. Finally we have analyzed the applicability of
hydrodynamics for the fireball evolution in heavy ion collisions
at various energies. Our analysis was based on a model
parametrization for the shear-viscosity and a rough estimate of
the relevant length scales governing the fireball expansion. Given
these assumptions we find that hydrodynamics should be applicable
for collisions at RHIC and LHC energies but not for SPS energies.

\vspace{0.25in}

\begin{acknowledgements}
The authors are grateful to Ulrich Heinz, Roy Lacey, Dirk Rischke,
Thomas Schaefer, Edward Shuryak, and Nu Xu for valuable
communications. The work is supported by the Director, Office of
Energy Research, Office of High Energy and Nuclear Physics,
Divisions of Nuclear Physics, of the U.S. Department of Energy
under Contract No. DE-AC02-05CH11231.
\end{acknowledgements}


\begin{thebibliography}{99}



\bibitem{Voloshin:2008dg}
  S.~A.~Voloshin, A.~M.~Poskanzer and R.~Snellings,
  arXiv:0809.2949 [nucl-ex].

\bibitem{hydro_review}
  U.~W.~Heinz,
  arXiv:0901.4355 [nucl-th].
   D.~A.~Teaney,
  arXiv:0905.2433 [nucl-th].
P.~Romatschke,
  arXiv:0902.3663 [hep-ph].



\bibitem{Schaefer:2009dj}
  T.~Schaefer and D.~Teaney,
 Rept.\ Prog.\ Phys.\  {\bf 72}, 126001 (2009).

\bibitem{Liao:2006ry}
  J.~Liao and E.~Shuryak,
  Phys.\ Rev.\  C {\bf 75}, 054907 (2007);
    Phys.\ Rev.\ Lett.\  {\bf 101}, 162302 (2008);
     Phys.\ Rev.\ Lett.\  {\bf 102}, 202302 (2009).
  M.~N.~Chernodub and V.~I.~Zakharov,
  Phys.\ Rev.\ Lett.\  {\bf 98}, 082002 (2007).


\bibitem{Carsten}
 Z.~Xu, C.~Greiner and H.~Stocker,
  Phys.\ Rev.\ Lett.\  {\bf 101}, 082302 (2008).
  J.~Noronha-Hostler, J.~Noronha and C.~Greiner,
Phys.\ Rev.\ Lett.\  {\bf 103}, 172302 (2009).

\bibitem{Pisarski}
R.~D.~Pisarski,
  Phys.\ Rev.\  D {\bf 74}, 121703 (2006).
  Y.~Hidaka and R.~D.~Pisarski,
  Phys.\ Rev.\  D {\bf 78}, 071501 (2008).



\bibitem{Liao:2009ni}
  J.~Liao and V.~Koch,
      Phys.\ Rev.\ Lett.\  {\bf 103}, 042302 (2009).


\bibitem{Koch:2009wk}
  V.~Koch,
  arXiv:0908.3176 [nucl-th].


\bibitem{Policastro:2001yc}
  G.~Policastro, D.~T.~Son and A.~O.~Starinets,
  Phys.\ Rev.\ Lett.\  {\bf 87}, 081601 (2001);
  P.~Kovtun, D.~T.~Son and A.~O.~Starinets,
  Phys.\ Rev.\ Lett.\  {\bf 94}, 111601 (2005).

\bibitem{Cherman:2007fj}
T.~D.~Cohen,
  Phys.\ Rev.\ Lett.\  {\bf 99}, 021602 (2007).
  A.~Cherman, T.~D.~Cohen and P.~M.~Hohler,
  JHEP {\bf 0802}, 026 (2008).



\bibitem{Csernai:2006zz}
  L.~P.~Csernai, J.~I.~Kapusta and L.~D.~McLerran,
  Phys.\ Rev.\ Lett.\  {\bf 97}, 152303 (2006).

\bibitem{Lacey:2006bc}
  R.~A.~Lacey {\it et al.},
  Phys.\ Rev.\ Lett.\  {\bf 98}, 092301 (2007).


\bibitem{Reynolds}
  A.~Bonasera and L.~P.~Csernai,
  Phys.\ Rev.\ Lett.\  {\bf 59}, 630 (1987);
  A. Bonasera, L.P. Csernai and B. Schurmann,
  Nucl.\ Phys.\ A {\bf 476}, 159 (1988).


\bibitem{Knudsen}
  C.~Gombeaud, T.~Lappi and J.~Y.~Ollitrault,
  Phys.\ Rev.\  C {\bf 79}, 054914 (2009).
  I.~Bouras {\it et al.},
  Phys.\ Rev.\ Lett.\  {\bf 103}, 032301 (2009).

\bibitem{Betz:2008me}
  B.~Betz, D.~Henkel and D.~H.~Rischke,
  Prog.\ Part.\ Nucl.\ Phys.\  {\bf 62}, 556 (2009).

\bibitem{Huang_book}
K.~Huang, {\it ``Statistical Mechanics''}, 2nd ed., Wiley, 1987.

\bibitem{Landau_book}
L.~D.~Landau and E.~M.~Lifshitz, {\it ``Fluid Mechanics''}, 2nd
ed., Butterworth-Heinemann, 1987.

\bibitem{Weinberg_book}
S.~Weinberg, {\it ``Gravitation and Cosmology: Principles and
Applications of the General Theory of Relativity''}, John Wiley \&
Sons, 1972.


\bibitem{Danielewicz:1984ww}
  P.~Danielewicz and M.~Gyulassy,
  Phys.\ Rev.\  D {\bf 31}, 53 (1985).


\bibitem{Schaefer:2009kj}
  T.~Schaefer,
  arXiv:0906.5399 [physics.flu-dyn].






\bibitem{supercritical}
E.~Ulrich Franck, J. Chem. Thermodynamics {\bf 19}, 225 (1987);
F.~Bencivenga {\it et al}, Europhys. Lett. {\bf 75}, 70 (2006);
F.~Gorelli {\it et al}, Phys. Rev. Lett. {\bf 97}, 245702 (2006);
V.~S.~Nikolayev {\it et al}, Phys. Rev. E {\bf 67}, 061202 (2003).



\bibitem{Meyer:2008dt}
  H.~B.~Meyer,
  Phys.\ Rev.\  D {\bf 79}, 011502 (2009).




\bibitem{NIST_webbook}
National Institute of Standards and Technology (NIST) Standard
Reference Database No. 69, June 2005, NIST Chemistry WebBook:
http://webbook.nist.gov/chemistry/. See, in particular,
Thermophysical Properties of Fluid Systems: High Accuracy Data for
a Select Group of Fluids.


\bibitem{Hirano:2005wx}
  T.~Hirano and M.~Gyulassy,
  Nucl.\ Phys.\  A {\bf 769}, 71 (2006).

\bibitem{Cheng:2007jq}
  M.~Cheng {\it et al.},
  Phys.\ Rev.\  D {\bf 77}, 014511 (2008).



\bibitem{fermi_atom_viscosity}
A.~Turlapov, et al, J. Low. Temp. Phys. {\bf 150}, 567 (2008).

\bibitem{Schaefer}
T.~Schaefer,   Phys.\ Rev.\  A {\bf 76}, 063618 (2007).

\bibitem{fermi_atom_sound}
J.~Joseph, et al, Phys.\ Rev.\ Lett. {\bf 98}, 170401 (2007).


\bibitem{Stephanov:2007fk}
  M.~A.~Stephanov,
  PoS {\bf LAT2006}, 024 (2006).

\bibitem{Fodor:2001pe}
  Z.~Fodor and S.~D.~Katz,
  JHEP {\bf 0203}, 014 (2002);
   JHEP {\bf 0404}, 050 (2004).

\bibitem{de Forcrand:2003hx}
  P.~de Forcrand and O.~Philipsen,
  Nucl.\ Phys.\  B {\bf 673}, 170 (2003).

\bibitem{Gavai:2004sd}
  R.~V.~Gavai and S.~Gupta,
  Phys.\ Rev.\  D {\bf 71}, 114014 (2005);
 Phys.\ Rev.\  D {\bf 72}, 054006 (2005).


\bibitem{Allton:2005gk}
  C.~R.~Allton {\it et al.},
  Phys.\ Rev.\  D {\bf 71}, 054508 (2005).

\bibitem{Philipsen:2005mj}
  O.~Philipsen,
  PoS {\bf LAT2005}, 016 (2006)
  [PoS {\bf JHW2005}, 012 (2006)].



\bibitem{Stephanov:1998dy}
  M.~A.~Stephanov, K.~Rajagopal and E.~V.~Shuryak,
  Phys.\ Rev.\ Lett.\  {\bf 81}, 4816 (1998);
  Phys.\ Rev.\  D {\bf 60}, 114028 (1999).

\bibitem{Lacey:2007na}
  R.~A.~Lacey, N.~N.~Ajitanand, J.~M.~Alexander, P.~Chung, J.~Jia, A.~Taranenko and P.~Danielewicz,
  arXiv:0708.3512 [nucl-ex].

\bibitem{Cheng:2008zh}
  M.~Cheng {\it et al.},
  Phys.\ Rev.\  D {\bf 79}, 074505 (2009).

\bibitem{Liao:2005pa}
  J.~Liao and E.~V.~Shuryak,
  Phys.\ Rev.\  D {\bf 73}, 014509 (2006);
 Nucl.\ Phys.\  A {\bf 775}, 224 (2006).

\bibitem{Kim:2009uu}
  K.~y.~Kim and J.~Liao,
  Nucl. Phys. B {\bf 822}, 201 (2009).



\bibitem{Koch:2008ia}
  V.~Koch,
  arXiv:0810.2520 [nucl-th].

\bibitem{Kestin:2008bh}
  G.~Kestin and U.~W.~Heinz,
  Eur.\ Phys.\ J.\  C {\bf 61}, 545 (2009).


\bibitem{BraunMunzinger:2001ip}
  P.~Braun-Munzinger, D.~Magestro, K.~Redlich and J.~Stachel,
  Phys.\ Lett.\  B {\bf 518}, 41 (2001).
  P.~Braun-Munzinger, K.~Redlich and J.~Stachel,
  arXiv:nucl-th/0304013.


\bibitem{Randrup:2009gp}
  J.~Randrup,
  Phys.\ Rev.\  C {\bf 79}, 054911 (2009).
J.~Randrup and J.~Cleymans,
  arXiv:0905.2824 [nucl-th].


\bibitem{Lublinsky:2007mm}
  M.~Lublinsky and E.~Shuryak,
  Phys.\ Rev.\  C {\bf 76}, 021901 (2007);
  Phys.\ Rev.\  D {\bf 80}, 065026 (2009).





\end{thebibliography}
\end{document}